\documentclass[twocolumn,preprintnumbers,amsmath,amssymb,prb]{revtex4}

\usepackage{graphicx}
\usepackage{dcolumn}
\usepackage{bm}

\begin{document}

\newcommand{\newc}{\newcommand}
\newc{\mbf}{\mathbf}
\newc{\boma}{\boldmath}
\newc{\beq}{\begin{equation}}
\newc{\eeq}{\end{equation}}
\newc{\beqar}{\begin{eqnarray}}
\newc{\eeqar}{\end{eqnarray}}
\newc{\beqa}{\begin{eqnarray*}}
\newc{\eeqa}{\end{eqnarray*}}

\newc{\bd}{\begin{displaymath}}
\newc{\ed}{\end{displaymath}}

\title{Orbits in a central force field: Bounded orbits}

\author{Subhankar Ray}
\email{subho@juphys.ernet.in}
\affiliation{Dept of Physics, Jadavpur University, Calcutta 700 032, India}
\author{J. Shamanna}
\affiliation{Physics Department, Visva Bharati University, Santiniketan 731235, India}

\date{August 1, 2003}

\begin{abstract}
The nature of boundedness of orbits of a particle
moving in a central force field is investigated.
General conditions for circular orbits and their stability 
are discussed. In a bounded central field orbit, a particle moves
clockwise or anticlockwise, depending on its angular momentum,
and at the same time oscillates between a minimum and a maximum
radial distance, defining an inner and an outer annulus. 
There are generic orbits suggested in popular texts
displaying the general features of a central orbit.
In this work it is demonstrated that some of these orbits,
seemingly possible at the first glance, 
are not compatible with a central force field. 
For power law forces, the general nature of boundedness
and geometric shape of orbits are investigated. 
\end{abstract}

\maketitle


\section{Introduction}

The central force motion is one of the oldest and widely studied 
problems in classical mechanics. Several familiar force-laws
in nature, e.g., Newton's law of gravitation, Coulomb's law,
van-der Waals force, Yukawa interaction, and Hooke's law are all 
examples of central forces.
The central force problem gives an opportunity to test one's 
understanding of the Lagrange's equation, Hamilton's equation, 
Hamilton Jacobi method, and classical perturbation. It also serves 
as an introduction to the concept of integrals of motion and 
conservation laws.
We need only to appeal to the principles of conservation of
energy and angular momentum to describe the nature and geometry
of the possible trajectories in central force motion.
Most books in classical mechanics 
\cite{goldstein,landau,symon,greenwood,sygr,sommerfeld},
treatise \cite{whitt,pars}, and advanced texts \cite{arnold,abmars}
discuss the central force problem.
In this article we present some interesting
features of bounded orbits in a central field.

\subsection{Kepler's Laws}
One of the most remarkable discoveries in the history of physics
is that of Keplerian orbits. 
A tremendous wealth of data on planetary positions
was collected by Tycho Brahe and Johannes Kepler
after detailed observation spread over several decades.
After a thorough analysis of this data
Johannes Kepler formulated three empirical 
laws that described and correlated the motion of the five planets 
then known: 
\begin{enumerate}
\item Each planet moves in an elliptical orbit, with the sun
at one of its foci.
\item The radius vector from the sun to each planet sweeps out 
equal areas in equal times.
\item The square of the periods ($T^2$) of the planets 
are proportional to the cube of the lengths of the 
corresponding semimajor axes ($a^3$).
\end{enumerate}

\subsection{Newtonian Synthesis}
Almost 100 years later Newton realized that the planets
go about in their nearly circular orbits around the sun under 
the influence of the same force that causes an apple to fall 
to the ground, i.e., gravitation. Newton's law of gravitation gave 
a theoretical basis to Kepler's laws. 
Kepler's laws can be derived from Newton's law of gravitation;
this is often referred to as the Newtonian synthesis.
Kepler's first and third laws are valid only in the specific case
of inverse square force. There are, however, certain general
features which are observed in all central field problems.
They include (i) certain conserved quantities (energy, and angular
momentum), (ii) planer nature of orbits, and (iii) constancy
of areal velocity (Kepler's second law).
A large class of central forces allows circular orbits (stable or
unstable), bounded orbits, and even closed and periodic orbits.
Certain common characteristics about the generic shapes of
bounded orbits can also be ascertained.

\section{Equations of motion and their first integrals}

\subsection{Central field orbits: confinement in a plane}
The central force motion between two bodies about their center 
of mass can be reduced to an equivalent one body problem in terms 
of their reduced mass $m$ and their relative radial distance $\mbf{r}$.
Hence in this reduced system, a body having the reduced mass
moves about a fixed center of force.

Consider the motion of a body under a central force, 
$\mbf{F} = \mbf{F}(\mbf{r}) = f(r) \hat{\mbf{r}}$ with the origin
as its force center.
The potential $V(r)$ from which this force is derived is also a 
function of $r$ alone, $\mbf{F} = - \mbox{\boma{$\nabla$}} 
V, \, V \equiv V(r)$.

On account of the central nature of the force, the mechanical
properties of the body do not vary under rotation in any manner
around the center of force.
Let the body be rotated through an infinitesimal angle
$\delta \mbox{\boma{$\theta$}}$, where the magnitude $\delta{\theta}$
is the angle of rotation while the direction is that of the axis
of rotation $\hat{\mbox{\boma{n}}}$. 
The change in the radius vector from the origin to
the body is $\mid \delta \mbf{r} \mid = r sin(\theta)
\delta{\theta}$, with $\delta \mbf{r}$ being perpendicular to $\mbf{r}$
and $\delta \mbox{\boma{$\theta$}}$. 
Hence $\delta \mbf{r} = \delta \mbox{\boma{$\theta$}}
\times \mbf{r}$. The change in velocity is similarly given by
$\delta \mbf{v} = \delta \mbox{\boma{$\theta$}}\times \mbf{v}$. 
The Lagrangian of the system is a function of $\mbf{r}$
and $\dot{\mbf{r}}$. The motion is governed by the Lagrange's equation,
\bd
\frac{d}{dt}\left( \frac{\partial L}{\partial \dot{\mbf{r}}} \right)- \frac{\partial L}{\partial \mbf{r}} = 0
\ed
If we now require that the Lagrangian of the system remain invariant 
under this rotation we obtain
\bd
\delta L = \frac{\partial L}{\partial \mbf{r}}\cdot \delta \mbf{r}+
\frac{\partial L}{\partial \dot{\mbf{r}}}\cdot \delta \dot{\mbf{r}} = 0 \; .
\ed
One can define generalized momentum as,
\bd
\mbf{p} = \frac{\partial L}{\partial \dot{\mbf{r}}}
\ed
From the Lagrange's equation we get,
\bd
\dot{\mbf{p}} \doteq \frac{d}{dt}\left( \frac{\partial L}{\partial \dot{\mbf{r}}} \right)= \frac{\partial L}{\partial \mbf{r}}
\ed
Replacing $\partial L/\partial \dot{\mbf{r}}$ by $\mbf{p}$
and $\partial L/\partial \mbf{r}$ by $\dot{\mbf{p}}$ in the
equation for $\delta L$ we get,
\bd
\dot{\mbf{p}}\cdot \delta \mbox{\boma{$\theta$}}\times \mbf{r}+
\mbf{p}\cdot \delta \mbox{\boma{$\theta$}}\times \dot{\mbf{r}} =0\; .
\ed
\bd
\delta \mbox{\boma{$\theta$}} \cdot \frac{d}{dt}(\mbf{r}\times \mbf{p}) =0 \; .
\ed
As $\delta \mbox{\boma{$\theta$}}$ is arbitrary we conclude 
$\mbf{l}=\mbf{r}\times \mbf{p}$ is a conserved quantity. $\mbf{l}$ is called
the angular momentum of the system. Since $\mbf{l}$ is a constant and is
perpendicular to $\mbf{r}$ it follows that the radius vector of
the particle lies in a plane perpendicular to $\mbf{l}$. This implies
that the motion of the particle in a central field is confined
to a plane.

\subsection{Lagrangian and equations of motion}

As the motion in a central force field is confined to a 
plane, it suffices to use plane polar coordinates. One 
may write the Lagrangian of the particle as,
\beq
L = \frac{1}{2}m \dot{\mbf{r}}^2-V(r)=\frac{1}{2}m(\dot{r}^2+r^2 \dot{\theta}^2)
-V(r) \; . 
\eeq
We assume the center of force to be at the origin.
The coordinates of the body of mass $m$ undergoing the central 
field motion are given by $(r,\theta)$.

The Lagrange's equation for the $\theta$ and $r$ coordinates are given 
respectively by,
\beqar
&&\frac{d}{d t} \left(\frac{\partial L}{\partial \dot{\theta}}\right) - \frac{\partial L}{\partial \theta} = 0 \\
&&\frac{d}{d t} \left(\frac{\partial L}{\partial \dot{r}}\right) - \frac{\partial L}{\partial r} = 0 
\eeqar

\subsection{First integrals and conservation laws}

The canonical momentum corresponding to $\theta$ is called the
angular momentum (or rather the magnitude of the angular momentum
that we discussed before),
\bd
p_{\theta} = \frac{\partial L}{\partial {\dot \theta}} = m r^2 \dot \theta = l
\ed
As $\theta$ is a cyclic coordinate, i.e., the Lagrangian is independent
of $\theta$, this angular momentum is conserved. This can be shown from
the Lagrange's equation for $\theta$.
\beqa
\frac{d}{d t}\left(\frac{\partial L}{\partial \dot{\theta}}\right) - \frac{\partial L}{\partial \theta} &=& 0 \\
\dot{p_{\theta}} = \frac{d}{d t}(m r^2 \dot{\theta}) &=& 0.
\eeqa
The corresponding integral of motion is,
\beq
m r^2 \dot{\theta} = l
\eeq 
and this $l$ can easily be shown to be the magnitude of the angular
momentum vector $\mbf{l} = \mbf{r} \times \mbf{p}$.

This conservation law is essentially equivalent to Kepler's 2nd Law:

\begin{figure}
\resizebox{3.4in}{!}
{\includegraphics*{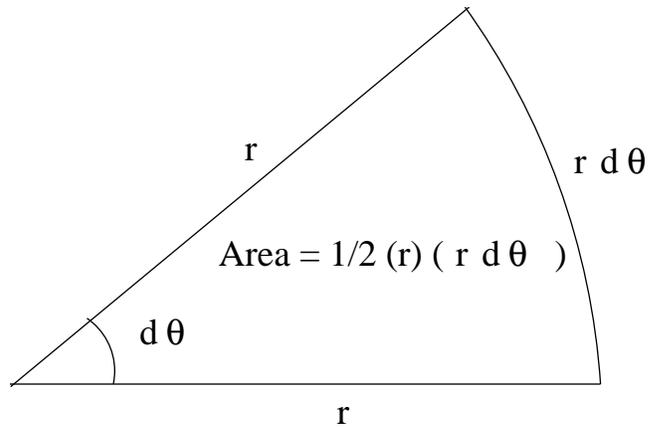}}
\caption{\label{fig:wide} Area swept by radius vector}
\end{figure}

The elementary triangular area swept by the radius vector in an infinitesimal
time interval $dt$ is,
\bd
dA = \frac{1}{2} r (r d\theta)
\ed
hence it follows from Eq. (4) that the rate of areal sweep is a constant.
\beq
\frac{dA}{dt} = \frac{1}{2} r^2 \dot{\theta} = \frac{l}{2m}
\eeq

There is another first integral of motion associated with the Lagrange's
equation for the $r$ coordinate.
\bd
\frac{d}{dt}(m \dot{r}) - m r {\dot{\theta}}^2 + \frac{\partial V}{\partial r}=0
\ed
The force in terms of the potential (conservative force) is given by, 
$f(r) = - \partial V/ \partial r$, whence the above equation becomes
\bd
m \ddot{r} - m r {\dot{\theta}}^2 = f(r)
\ed
Using the first integral of motion, one can convert this second equation
into an equation for $r$ alone.
\bd
m \ddot{r} = - \frac{d}{dr} \left( V + \frac{1}{2} \frac{l^2}{m r^2} \right)
\ed
Integrating we get 
\beq 
\frac{1}{2} m {\dot{r}}^2 + \frac{1}{2} \frac{l^2}{m r^2} + V(r) = E 
\eeq
where $E$ is a constant of integration, called the energy.
This is the law of conservation of total mechanical energy.

It is interesting to note that for motion in a general force field,
$l=m r^2 \dot{\theta}$ remains invariant even though $r$ and $\dot{\theta}$
vary with time (see Fig. 2). Similarly 
$ E= m {\dot{r}}^2/2 + l^2/(2 m r^2) + V(r)$
remains a constant even though $\dot{r}$ and $r$ and hence
$m \dot{r}^2/2$ and $V(r) + l^2/(2m r^2)$ each varies with time
(see Fig. 3).

\begin{figure}
\resizebox{3.4in}{!}
{\includegraphics*{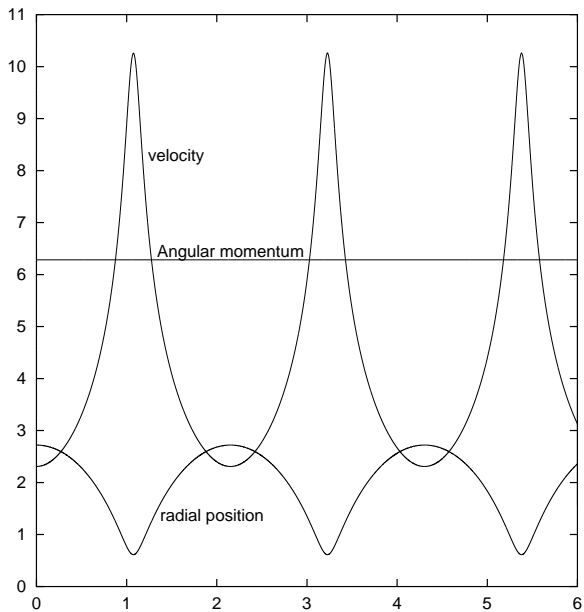}}
\caption{\label{fig:wide} Constancy of angular momentum in an inverse square potential. Total energy is $60\%$ of the minimum of effective potential $\tilde{V}(r)$.}
\end{figure}

\begin{figure}
\resizebox{3.4in}{!}
{\includegraphics*{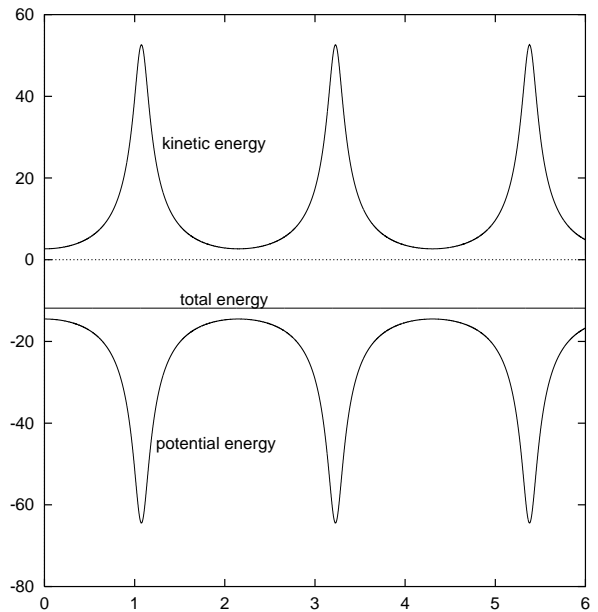}}
\caption{\label{fig:wide} Constancy of total energy in an inverse 
square potential. Total energy is $60\%$ of the minimum of effective 
potential $\tilde{V}(r)$.}
\end{figure}

Lagrange's equations are two second order Ordinary Differential
Equations (ODE) in $r$ and $\theta$. However they decouple, i.e.,
each equation is expressible in terms of either $r$ or $\theta$.
On integrating each equation once we get
the first integrals of motion namely, the total mechanical energy and
the angular momentum.
A further integration will yield the complete solution to the problem.
This second integration introduces two more constants of integration
namely, the initial radial ($r_0$) and angular ($\theta_0$) positions.

\subsection{Equation for the orbit}
From the equation giving energy as a first integral of motion, we get 
the expression for radial velocity,
\beq
\dot{r} = \sqrt{\frac{2}{m}\left(E-V-\frac{l^2}{2mr^2}\right)}
\eeq
On integration we get,
\beq
t = \int_{r_0}^{r} \frac{dr}{\sqrt{2/m(E-V-l^2/(2mr^2))}}.
\eeq
This relation can be inverted to give $r$ as a function of $t$, $r = r(t)$.
The other first integral gives,
\bd
\dot{\theta} = \frac{l}{mr^2}.
\ed
Substituting the expression for $r$ as a function of $t$, 
and integrating,
\beq
\theta - \theta_0 = \frac{l}{m} \int_0^{t} \frac{dt}{r(t)^2}.
\eeq
These two expressions for $r(t)$ and $\theta(t)$ express
the equation of the orbit for the particle in a central 
field in terms of time $t$ as a parameter.

Instead of expressing the orbit parametrically in terms of $t$,
one often wants to express the orbit directly as an equation 
connecting $r$ and $\theta$. 
Such an equation may be obtained by eliminating $t$ from the
above expressions for $\dot{r}$ and $\dot{\theta}$.
\bd
\frac{d \theta}{d r} = \frac{l/(mr^2)}{\sqrt{2/m(E-V(r) - l^2/(2mr^2))}}
\ed
On integration this yields,
\bd
\theta - \theta_0 = \int_{r_0}^{r} \frac{l/(mr^2)}{\sqrt{2/m(E-V(r) - l^2/(2mr^2))}} dr
\ed

For understanding the qualitative nature of motion in a central field 
one looks at the equivalent one dimensional problem.
\beqa 
\dot{r} &=& \sqrt{\frac{2}{m} (E-V(r)-\frac{l^2}{2mr^2})} \\
&=& \sqrt{\frac{2}{m} (E-\tilde{V}(r))}.
\eeqa
We call $\tilde{V}$ the effective potential, introduced to make the 
problem similar to that of a particle moving in a one 
dimensional potential field. The effective radial force ($\tilde{f}(r)$)
is connected to the effective radial potential ($\tilde{V}(r)$) by the 
expected relation,
\beq
\tilde{f}(r) = - \frac{\partial \tilde{V}(r)}{\partial r}
\eeq
At a point where the effective potential $\tilde{V}(r)$ 
equals the energy $E$, the radial velocity vanishes ($\dot{r} =0$).
In one dimensional motion this corresponds to a particle 
coming momentarily to rest, and having zero kinetic energy.
However in the case of central field, the motion is not really
one dimensional, and even for $\dot{r} =0$,
the particle is not at rest 
($\mbf{v}=r \dot{\theta} \hat{\mbox{\boma{$\theta$}}}$), 
and it has a non-zero kinetic energy ($(1/2 m) r^2 \dot{\theta}^2$). 

For the inverse square force, as in the case of gravitation,
we have $f(r) = -k/r^2$ and $V(r) = -k/r$.
Effective potential (see Fig. 4) is given by, 
\beq
\tilde{V}(r) = - \frac{k}{r} + \frac{l^2}{2mr^2}
\eeq

\begin{figure}
\resizebox{3.4in}{!}
{\includegraphics*{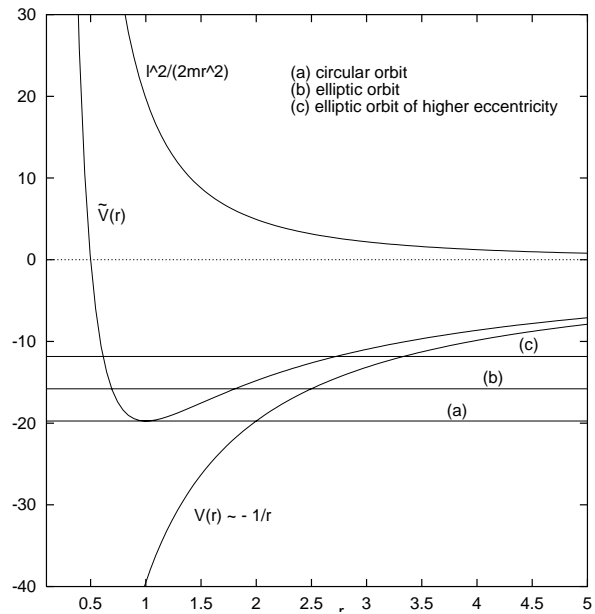}}
\caption{\label{fig:wide} Effective potential for an inverse square force}
\end{figure}

The following properties of the effective potential are easily noted:
\begin{enumerate}
\item $\tilde{V}(r) = 1/r^2 (l^2/2m-kr)$, 
as $r\rightarrow0$ the term within bracket is essentially $l^2/(2m)$
and hence $\lim_{r\rightarrow0}\tilde{V}(r) \rightarrow +\infty$
\item $\lim_{r\rightarrow \infty} \tilde{V}(r) = 0$
\item $\tilde{V}(r)=-1/r(k-l^2/2mr^2)$ and for large values of $r$,
$l^2/2mr^2$ is negligible compared to $k$, hence 
$\tilde{V}(r)$ has a negative value.
\item At $r^* = l^2/2mk$ the function $\tilde{V}(r)$ intersects the $r$ axis,
i.e., $\tilde{V}(r^*) = 0$.
\item $\tilde{V}(r)$ reaches a minimum at $r_0 = 2 \cdot r^* =l^2/mk$.
With $\partial \tilde{V}/\partial r |_{r_0} = 0$ and
$\partial^2 \tilde{V}/\partial^2 r |_{r_0} > 0$.
\end{enumerate}

For total energy $E=\tilde{V}(r_0)$ the particle has zero radial
velocity at $r=r_0$, and no other radial position is physically
accessible, since $\dot{r}$ becomes imaginary at $r \neq r_0$.
This corresponds to circular motion. 

For $0 > E > \tilde{V}(r_0)$ there is a range of radial 
positions ($r_{max} \geq r \geq r_{min}$) for which $E \geq \tilde{V}(r)$.
The particle can move in this range of $r$ with varying $\dot{r}$.
The radial velocity $\dot{r}$ vanishes at the end points
$r_{max}$ and $r_{min}$ where the energy $E$ equals the effective
potential $\tilde{V}$ and $\dot{r}$ reaches a maximum at $r=r_0$
The points $r_{max}$ and $r_{min}$ are called the turning points.
The central field particle cannot move beyond these points, 
as the energy becomes less than the effective potential,
and the expression for radial velocity ($\dot{r}$) turns imaginary.

$E < \tilde{V}(r_0)$ is a physically impossible situation,
since no (radial) position is physically allowed for the particle.

For $E \geq 0$ we have an unbounded motion, where 
the particle can fly off to infinity.
Thus in the case of inverse square force field we can have
bounded ($E < 0$) or unbounded ($ E \geq 0$) motion depending 
on the energy of the particle. In particular for $E = 0$ it is 
parabolic and for $E > 0$ it is hyperbolic.

\section{Existence and stability of circular orbits for central forces}

At all positions other than where the effective potential is
a minimum or a maximum, we have a net effective force 
$\tilde{f}(r) = - \partial \tilde{V}/\partial r \neq 0$. 
When the total energy $E$ is not equal to the minimum of 
effective potential $\tilde{V}(r)$, at the points of 
instantaneous zero radial velocity the particle is pushed away.
If the effective potential has a minimum and energy is greater
than that minimum then the radial distance has a lower and an 
upper bound and the particle moves between these radial limits.
On the other hand if the effective potential has a maximum, and the energy is
less than that maximum, the effective force pushes the particle 
away from the positions of zero radial velocity in an inward or
an outward spiral.
If the total energy is equal to the maximum or minimum of the effective 
potential, the system can stay with zero radial velocity, and 
hence move in a circular orbit.

\begin{figure}
\resizebox{3.4in}{!}
{\includegraphics*{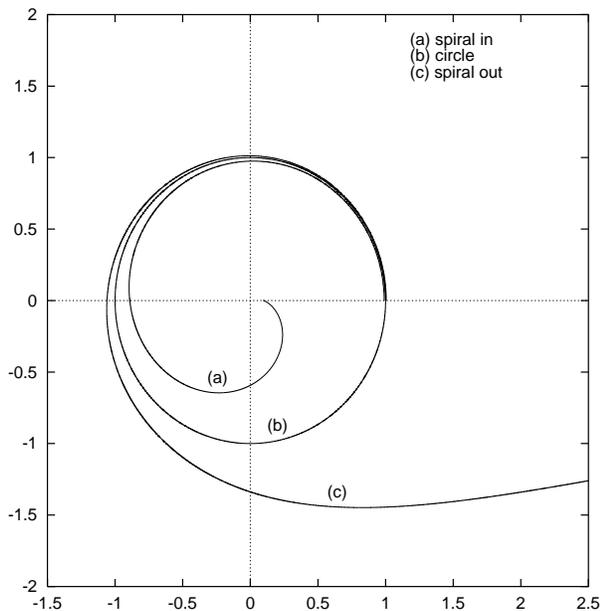}}
\caption{\label{fig:wide} Spiralling orbits, $f(r) \sim - 1/r^4 $}
\end{figure}

The condition for circular orbit is,
\beq
\left. \frac{\partial \tilde{V}}{\partial r} \right|_{r_0} = \left.\frac{\partial V}{\partial r}\right|_{r_0} -\frac{l^2}{mr_0^3} = 0
\eeq
whence we get,
\beq
f(r_0) = - \left.\frac{\partial V}{\partial r}\right|_{r_0} = - \frac{l^2}{mr_0^3}
\eeq

The negative sign on the right hand side clearly shows
that the force must be attractive. 
The particle moves in a circular orbit,
since the force of attraction due to the central field 
provides the necessary centripetal force.

For a given attractive central force ( $f(r)$ and $V(r)$ given )
it is possible to have a circular orbit of radius $r_0$ provided
the angular momentum and energy of the particle are given by,
\beqar
l^2 &=& - m r_0^3 f(r_0) \\
E &=& V(r_0) + \frac{l^2}{2 m r_0^2}
\eeqar

However when the effective potential has a maximum the system 
is in an unstable circular orbit.
A small deviation from this radial position causes the orbit 
to be unbounded. The effective force that comes into play
makes the particle move away from the position of zero radial 
velocity in an inward or outward spiral (Fig. 5). 
When the effective potential has a minimum, the 
effective forces cause the particle to remain in a bound 
orbit confined in an annular space. For small deviations
the annular radii are nearly the same, and hence the orbit
is close to a circle.
One can understand the stability question by 
studying the forces (restoring or unsettling) that come into
play when the system is moved infinitesimally from the position 
of circular orbit. 

The circular orbit is stable if,
\beq
\frac{\partial^2 \tilde{V}}{\partial r^2} > 0.
\eeq
That is equivalent to,
\beqa
&& \frac{{\partial}^2 \tilde{V}}{\partial r^2} = \left.\left(-\frac{\partial f}{\partial r} + \frac{3 l^2}{m r^4} \right)\right|_{r_0} >0 \\
&& \frac{\partial f}{\partial r}|_{r_0} < \frac{3 l^2}{m {r_0}^4} 
\eeqa
Using Eq. (13),
\beq
\frac{\partial f}{\partial r}|_{r_0} < - \frac{3 f(r_0)}{r_0}
\eeq

\section{Nature of bounded orbits}

General bounded motion has both lower and upper bounds.
It means that the particle cannot approach nearer than some
minimum or move farther than some maximum distance.
One has to remember that the angular velocity has a constant 
sign, same as that of the constant angular momentum,
throughout the motion. However its magnitude decreases with
increase in the radial distance ($\sim 1/r^2$). Together with
the angular motion, the radial distance changes from 
$r_{max}$ to a $r_{min}$, then
back to $r_{max}$ and so on. From this general nature of motion
two families of generic orbits are suggested in popular texts \cite{goldfig}
(Fig. 6, Fig. 7). However it can be shown that the generic
types shown in the later figure (Fig. 7) are not feasible for any attractive 
potential. 

\begin{figure}
\resizebox{!}{3.4in}
{\includegraphics*{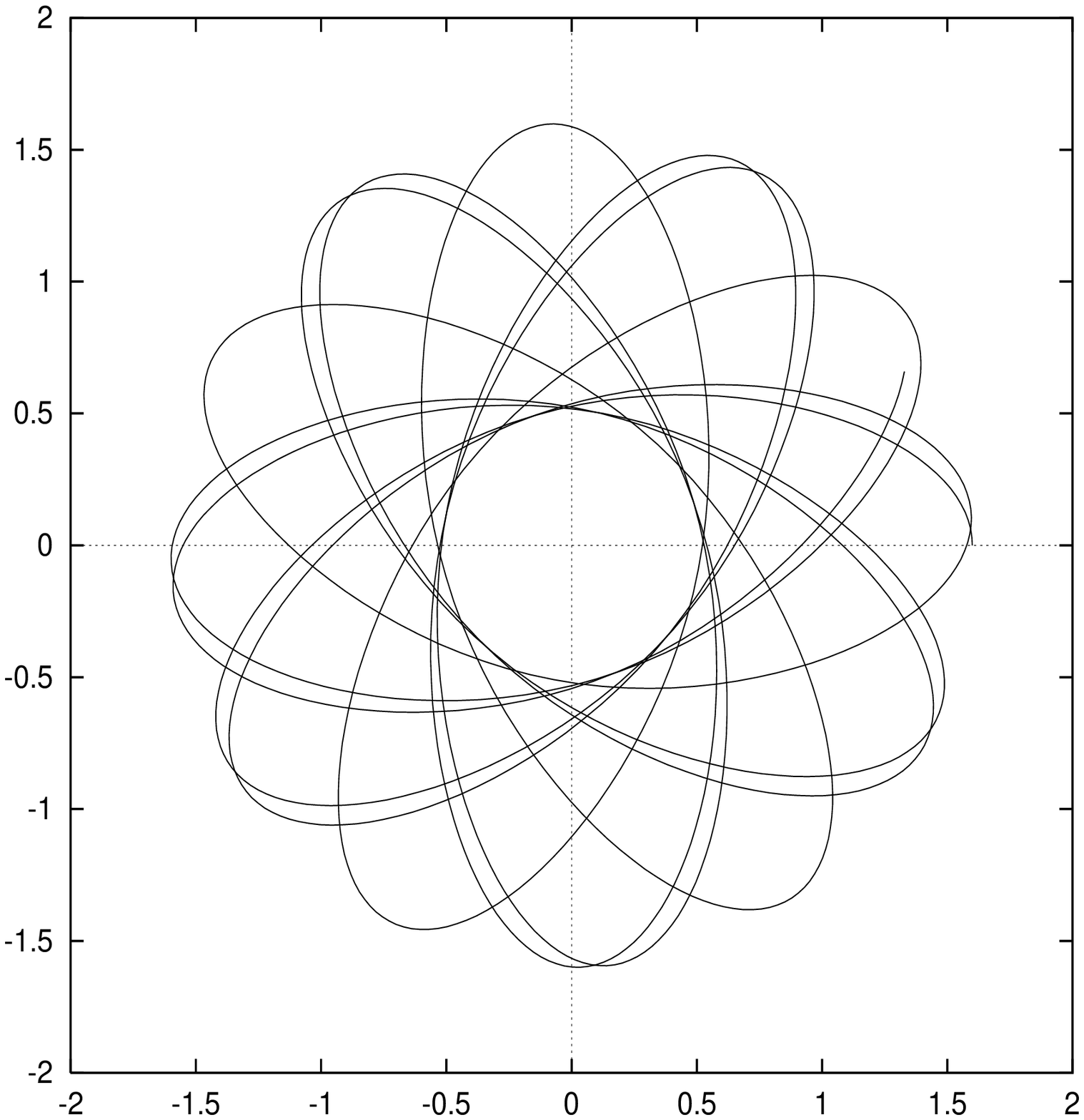}}
\resizebox{!}{3.4in}
{\includegraphics*{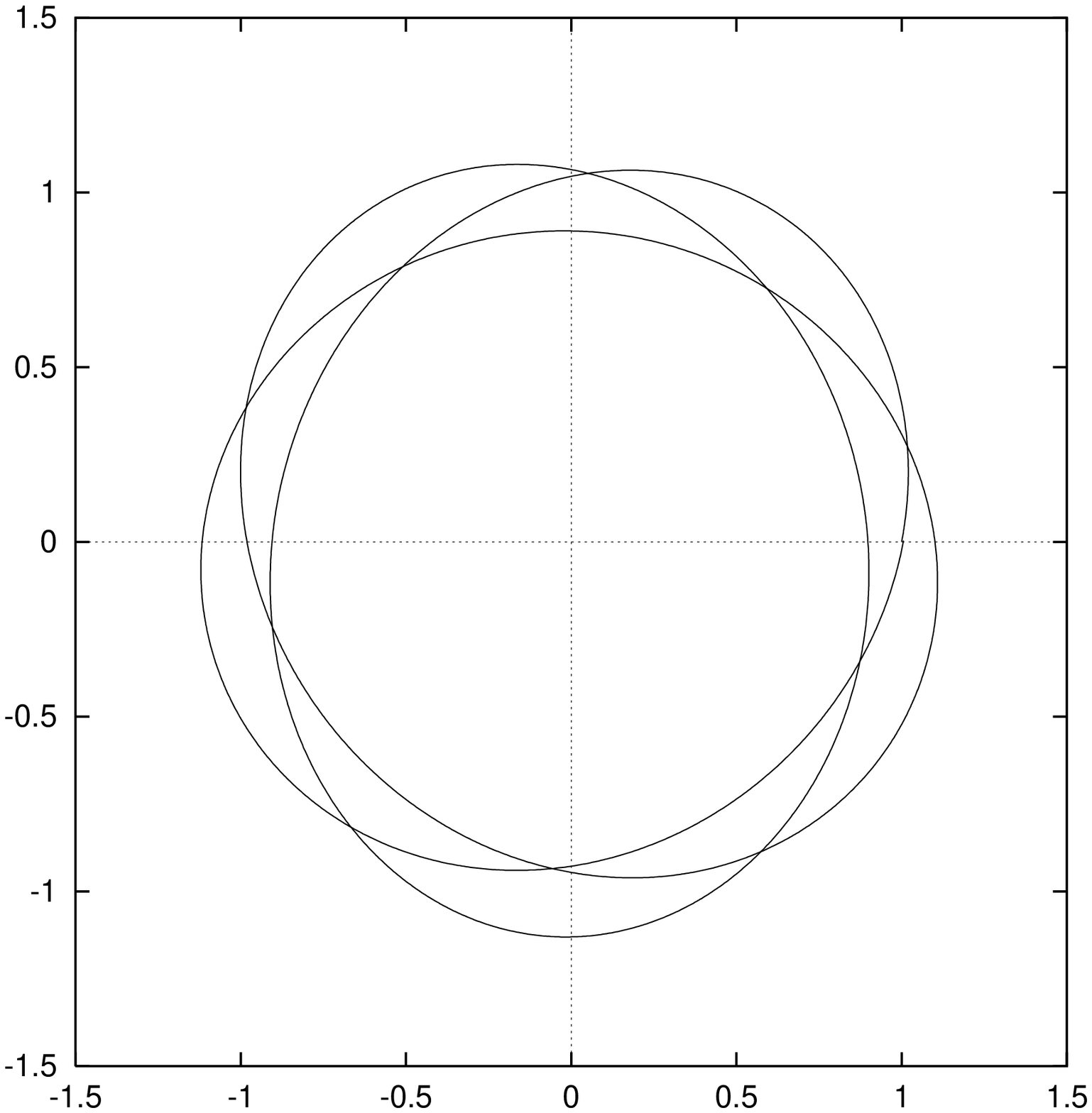}}
\caption{Generic orbits in central field (allowed)}
\end{figure}

\begin{figure}
\resizebox{!}{3.4in}
{\includegraphics*{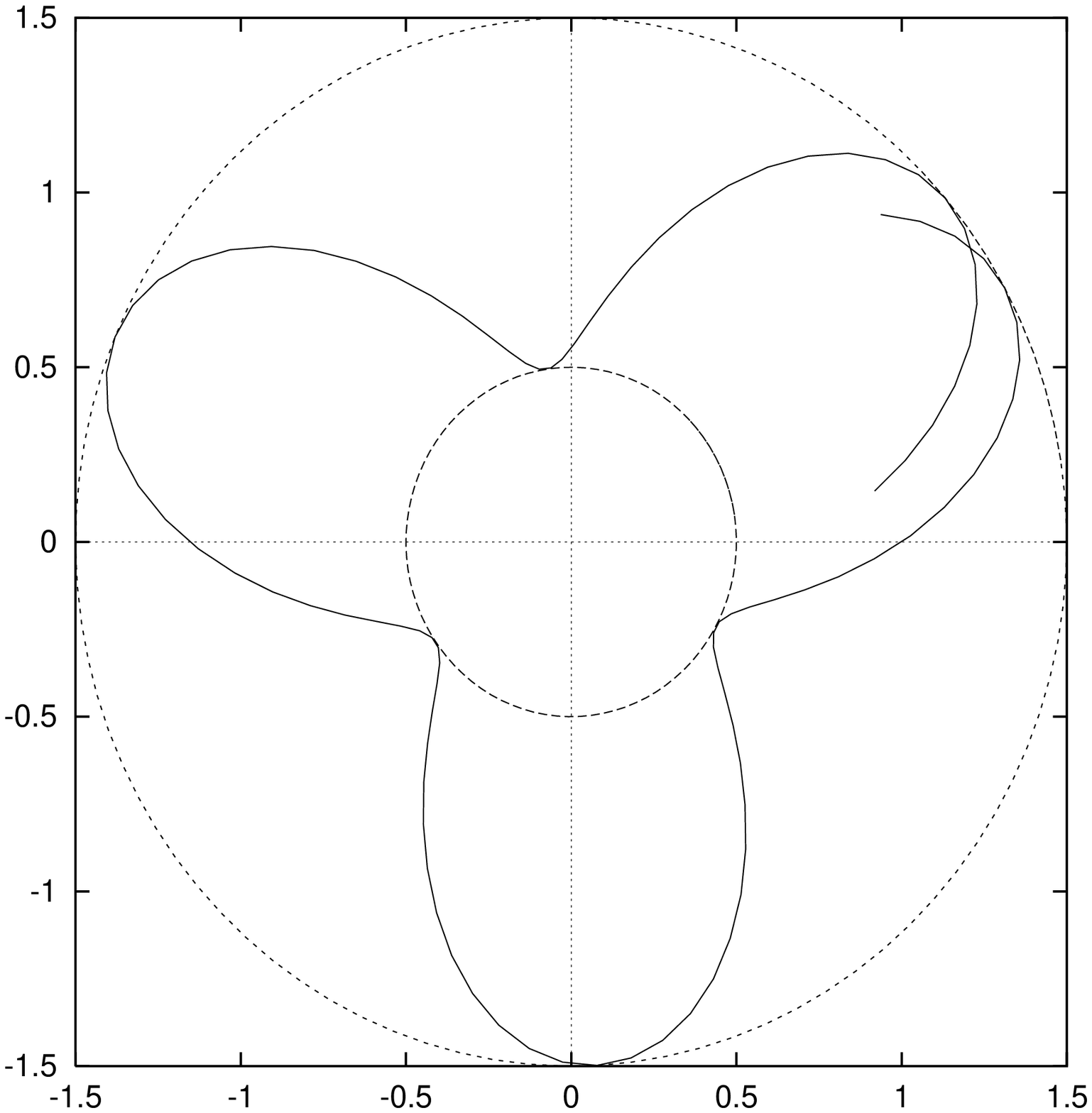}}
\resizebox{!}{3.4in}
{\includegraphics*{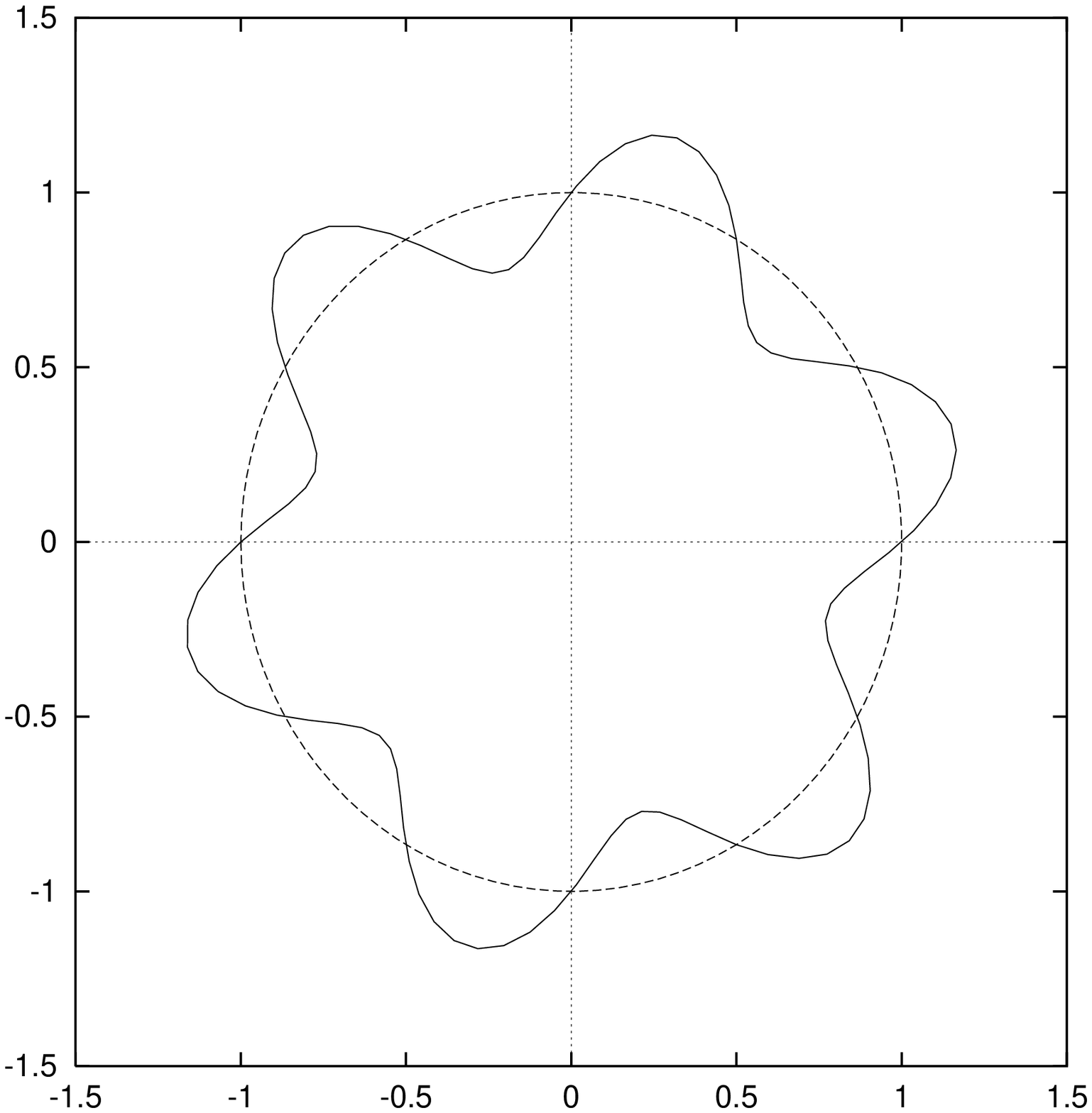}}
\caption{Generic orbits in central field (not allowed)}
\end{figure}

We consider below a general bounded orbit confined in an annular region.
We would like to investigate the nature of the orbit
close to the point where it touches the inner or
the outer annulus.
Let us choose the reference line (polar line) of the coordinate system
such that the orbit touches the annular ring at $\theta = 0$, or more 
specifically at $r=R, \theta=0$.
Since at this point $P (r=R,\theta=0)$ the orbit is at its closest
(or farthest) approach from the pole, $(\partial r/ \partial t)_P = 0$.
As $r$ is a function of $\theta$, using Eq. (4) we get for motion along
the trajectory,
\bd
\frac{dr}{dt} = \frac{l}{mr^2} \frac{dr}{d \theta}
\ed
hence
\beq
r'(0) = \left.\frac{dr}{d\theta}\right|_{\theta=0} = 0
\eeq
The equation of motion for $r$ is given by,
\bd
m \ddot{r} - m r {\dot{\theta}}^2 = f(r) 
\ed
From which we get $r''(\theta)$ along the trajectory,
\beqa
m \frac{l}{mr^2} \frac{d}{d \theta}\left(\frac{l}{mr^2} \frac{dr}{d \theta}\right) -mr \frac{l^2}{mr^3} &=& f(r) \\
\frac{d^2 r}{d {\theta}^2} &=& \frac{mr^4}{l^2} \tilde{f}(r) \\
&=& r + \frac{mr^4}{l^2} f(r)
\eeqa
One can expand $r(\theta)$ in a Taylor series 
in $\theta$, and remember that $r'(0) =0$,
\beqa
r(\theta)_{traj} &=& r(0)+r'(0) \theta + r''(0) \frac{\theta}{2!} + h.o.\\
&=& r(0) + \frac{mr(0)^4}{l^2} \tilde{f}(r(0))  \frac{{\theta}^2}{2!} + h.o. \\
&=& r(0) + \left( r(0) + \frac{mr(0)^4}{l^2} f(r(0)) \right) \frac{{\theta}^2}{2!} + h.o.
\eeqa
Consider a tangent to the annulus at the point $P$, Fig. 8,
\bd
\frac{r(0)}{r(\theta)} = \cos (\theta)
\ed
which can be expanded to give $r(\theta)_{tan}$ along the tangent,
\bd
r(\theta)_{tan} = r(0) + r(0) \frac{{\theta}^2}{2} + h.o.
\ed
Notice that the constant or the $\theta$ independent terms
in $r(\theta)_{traj}$ and $r(\theta)_{tan}$ 
are equal, and
the first leading power of $\theta$ is $\theta^2$ in both the cases.
The coefficient of $\theta^2$ (curvature) will determine the nature of the
trajectory in reference to the tangent.

\begin{figure}
\resizebox{3.4in}{!}
{\includegraphics{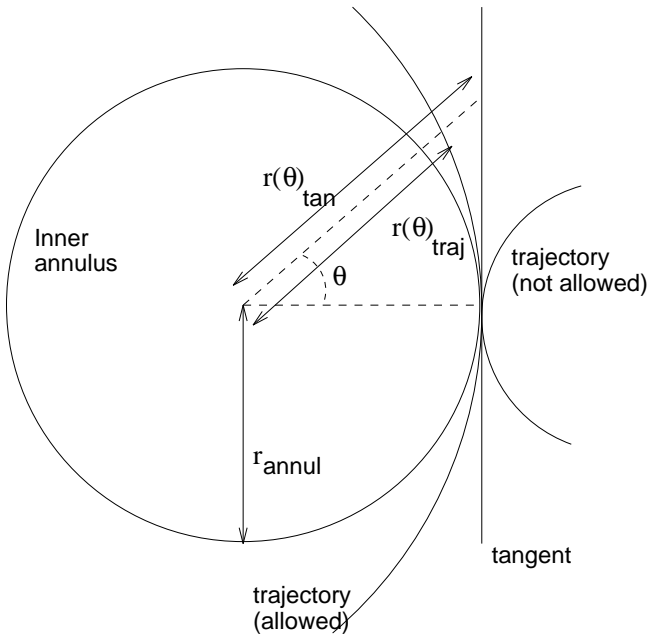}}
\caption{Curvature of the trajectory in comparison to the tangent to the
inner annulus}
\end{figure}

\begin{figure}
\resizebox{3.4in}{!}
{\includegraphics{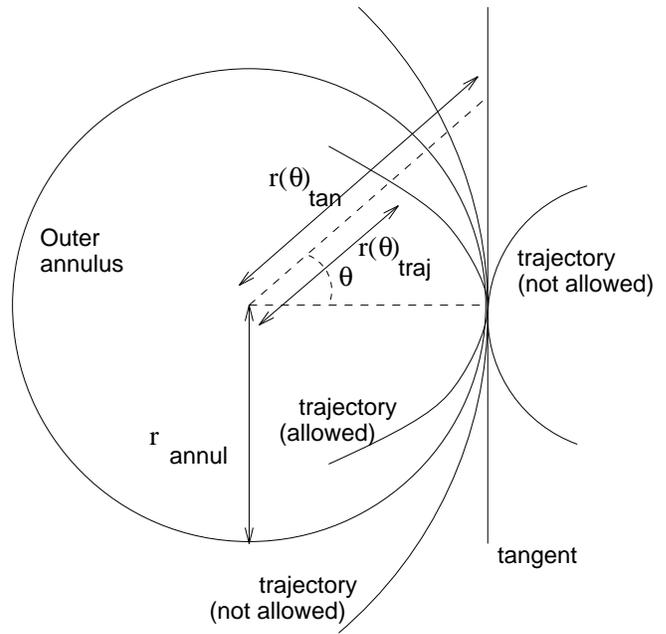}}
\caption{Curvature of the trajectory in comparison to the outer annulus}
\end{figure}

We first study the curvature of the orbit in reference to the
tangent. The force being attractive $f(r(0)) < 0 $,
for small angular distance ($\Delta(\theta)$) away from the point $P$,
\beq
r( \Delta \theta)_{traj} < r( \Delta \theta)_{tan}
\eeq
which means that the orbit bends more sharply than the tangent
and stays closer to the inner annulus for nearby points (Fig. 8).
Hence the second type of orbits shown in some texts (Fig. 7) 
are not possible for central forces.
For the outer annulus the analysis with respect to tangent is
not very meaningful as one can set up an even stronger bound for
its orbit, directly considering the potential $\tilde{V}(r)$.
In this case the effective potential has a positive slope with
respect to the radius, and hence a negative effective force
($\tilde{f}(r) < 0$). The orbit should not only remain nearer 
than the tangent, but even nearer than the outer annulus.
This condition is confirmed if we study the orbit near the point
$P$ where it touches the outer annulus.
\bd
r(\Delta \theta)_{traj} = r(0) + \frac{mr(0)^4}{l^2} \tilde{f}(r) \frac{(\Delta \theta)^2}{2!} + h.o. 
\ed
hence
\beq
r(\Delta \theta)_{traj} < r(0) = r_{max}
\eeq 
This confirms that the outer annulus is indeed the outer 
bound of the trajectory.
The nature of the orbit near the point of contact $P$
at the outer annulus is shown in figure 9.

\section{Bounded orbits for the power law central force}

\subsection{Existence of stable circular orbit}

For the case of a power law central potential
\beq
V(r) = \frac{k}{r^n}, \;\;\;\;\;\;\;\; f(r) = -\frac{nk}{r^{n+1}}
\eeq
From the stability condition Eq. (17),
\bd
\frac{\partial f}{\partial r}|_{r_0}  < - \frac{3 f(r_0)}{r_0}
\ed
we find,
\beqar
\frac{(n+1)nk}{r_0^{n+2}} &<& - 3 \left( - \frac{k}{r_0^{n+1}} \right) \frac{1}{r_0} \nonumber \\
n &<& 2
\eeqar
Hence the circular orbit is stable for an attractive power
law potential that varies slower than inverse square 
(or a force that varies slower than inverse cube).

\subsection{\label{sec:level2} Study of boundedness of orbits}

\subsubsection{$V(r) = -a/r^n$ where $n>2$}

\begin{figure}
\resizebox{3.4in}{!}
{\includegraphics*{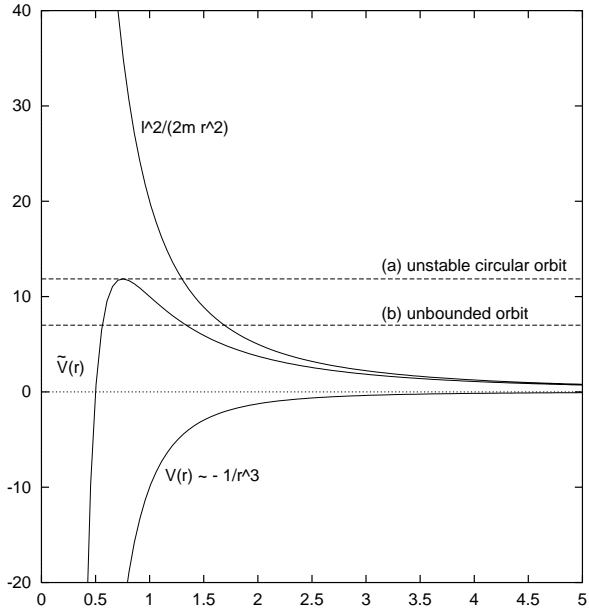}}
\caption{\label{fig:wide} Effective potential for $f(r) \sim -1/r^4$ force}
\end{figure}

Consider the case when $ n > 2 $.
We have the effective potential,
\beq
\tilde{V}(r) = - \frac{a}{r^n} + \frac{l^2}{2mr^2}
\eeq
\begin{enumerate}
\item $\tilde{V}(r) = -1/r^n(a-l^2 \cdot r^{n-2}/(2m))$, and as 
$r \rightarrow 0$ we may neglect $l^2 \cdot r^{n-2}/(2m)$ in comparison 
to $a$, thus $\lim_{r \rightarrow 0}\tilde{V}(r) = - \infty$.
\item $\lim_{r \rightarrow \infty} \tilde{V}(r) = 0$.
\item For large but finite $r$, 
$\tilde{V}(r) = -1/r^2(a/r^{n-2} - l^2/(2m))$, and $a/r^{n-2}$ is
negligible compared to $l^2/(2m)$, and hence $\tilde{V}(r)$ is positive.
\item $\tilde{V}(r)$ intersects the $r$ axis at $r^*$ where 
$(r^*)^{n-2}=2am/l^2$.
\item $\tilde{V}(r)$ has a maximum at $r_0$ where $r_0^{n-2}= anm/l^2$.

For a maximum point of $\tilde{V}$ we should have 
$\partial \tilde{V}/\partial r \mid_{r_0} =0$
and $\partial ^2 \tilde{V}/\partial r^2 \mid_{r_0} < 0$.
\bd
\frac{\partial \tilde{V}}{\partial r} = \frac{an}{r^{n+1}} - 
\frac{l^2}{mr^3} =0 \; .
\ed
For the point of extremum
\bd
{r_0}^{n-2} = \frac{amn}{l^2} \;.
\ed
Finding the second derivative of $\tilde{V}$ with respect to $r$
at the point $r_0$,
\beqa
\frac{d^2\tilde{V}}{dr^2} &=& -\frac{an(n+1)}{r^{n+2}} + \frac{3l^2}{mr^4} \\
&=& \frac{1}{r^4} \left(-\frac{an(n+1)}{r^{n-2}} + \frac{3 l^2}{m}\right) \\
&=&\frac{1}{r_0^4} \left(-(n+1) \frac{l^2}{m} + \frac{3l^2}{m}\right) \\
&=& \frac{1}{r_0^4}\frac{l^2}{m}(2-n) < 0
\eeqa
\end{enumerate}

Hence it is a point of maximum.
For $E > \tilde{V}(r_0)$ we always have an unbounded orbit.
For $E < \tilde{V}(r_0)$ the orbit is semibounded, bounded above
or bounded below, according to its initial state. The particle
either spirals in or spirals out. For $E=\tilde{V}(r_0)$ we
get an unstable circular orbit.

\subsubsection{$V(r) = -a/r^n$ where $n = 2$}
The effective radial potential is
$\tilde{V}(r) = - a/r^2 + l^2/(2mr^2)$.
This is essentially an attractive or repulsive
inverse square term.
\beqa
\tilde{V}(r) = - \frac{\tilde{a}}{r^2}, \;\;\; && \tilde{a}>0 \;\;\; if \;\;\; a > \frac{l^2}{2m} \\
&& \tilde{a} <0 \;\;\; if \;\;\; a < \frac{l^2}{2m}
\eeqa
This potential cannot give circular orbit ever.
If the effective potential is attractive and $E<0$ it
has an upper bound of radial distance. A typical orbit
would therefore be an inward spiral. For a repulsive effective 
potential we will have outward spiral moving to infinite
radial distance.

\begin{figure}
\resizebox{3.4in}{!}
{\includegraphics*{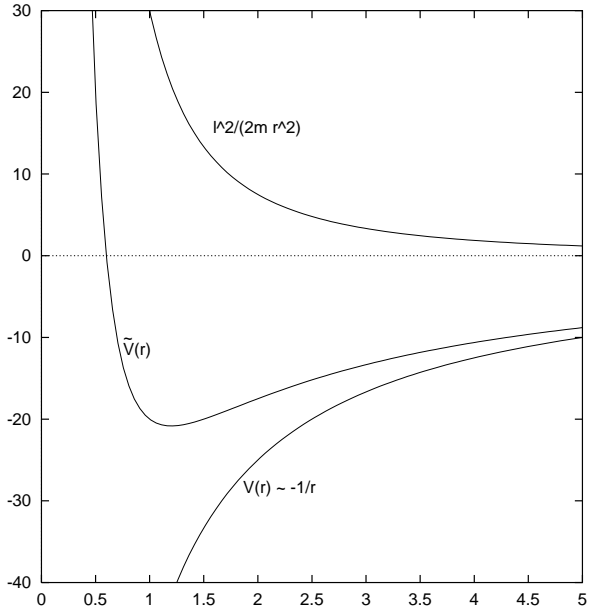}}
\caption{\label{fig:wide} Effective potential for $f(r)\sim - 1/r^2$ force}
\end{figure}

\subsubsection{$V(r) = -a/r^n$ where $2>n>0$}
In this case the effective potential is,
\beq
\tilde{V}(r) = - \frac{a}{r^n} + \frac{l^2}{2mr^2}
\eeq

We have the following properties 
of $\tilde{V}(r)$

\begin{enumerate}
\item  $\tilde{V}(r) = 1/r^2 (l^2/2m - a/r^{n-2})$ and as $r \rightarrow0$
we may neglect $a/r^{n-2}$ in comparison to $l^2/2m$, and thus
$\lim_{r \rightarrow 0} \tilde{V}(r) \rightarrow +\infty$.
\item $\lim_{r \rightarrow \infty} \tilde{V}(r) = 0$.
\item $\tilde{V}(r)=-1/r^n (a-l^2/2mr^{2-n})$ and for large but finite
values of $r$,  $l^2/2mr^{2-n}$ is negligible compared to $a$, hence 
$\tilde{V}(r)$ is negative.
\item $\tilde{V}(r)$ intersects the $r$ axis at $r^*$ where 
$(r^*)^{2-n}=l^2/(2am)$.
\item $\tilde{V}(r)$ reaches a minimum at $r_0$
where ${r_0}^{2-n}=l^2/amn$. \\
For a minimum point of $\tilde{V}$ we should have 
$\partial \tilde{V}/\partial r \mid_{r_0} =0$
and $\partial ^2 \tilde{V}/\partial r^2 \mid_{r_0} > 0$.
\bd
\frac{\partial \tilde{V}}{\partial r} = \frac{an}{r^{n+1}} - 
\frac{l^2}{mr^3} =0 \; .
\ed
For the point of extremum
\bd
{r_0}^{2-n} = \frac{l^2}{amn} \;.
\ed
Finding the second derivative of $\tilde{V}$ with respect to $r$
at the point $r_0$,
\beqa
\frac{d^2\tilde{V}}{dr^2} &=& -\frac{an(n+1)}{r^{n+2}} + \frac{3l^2}{mr^4} \\
&=& \frac{1}{r^4} \left(-\frac{an(n+1)}{r^{n-2}} + \frac{3 l^2}{m} \right) \\
&=&\frac{1}{r_0^4} \left( -(n+1) \frac{l^2}{m} + \frac{3l^2}{m} \right) \\
&=& \frac{1}{r_0^4}\frac{l^2}{m}(2-n) > 0
\eeqa
\end{enumerate}

Hence it is a point of minimum.
For this case the orbit can be bounded or unbounded 
depending on the total energy $E$ of the particle.

\begin{figure}
\resizebox{3.4in}{!}
{\includegraphics*{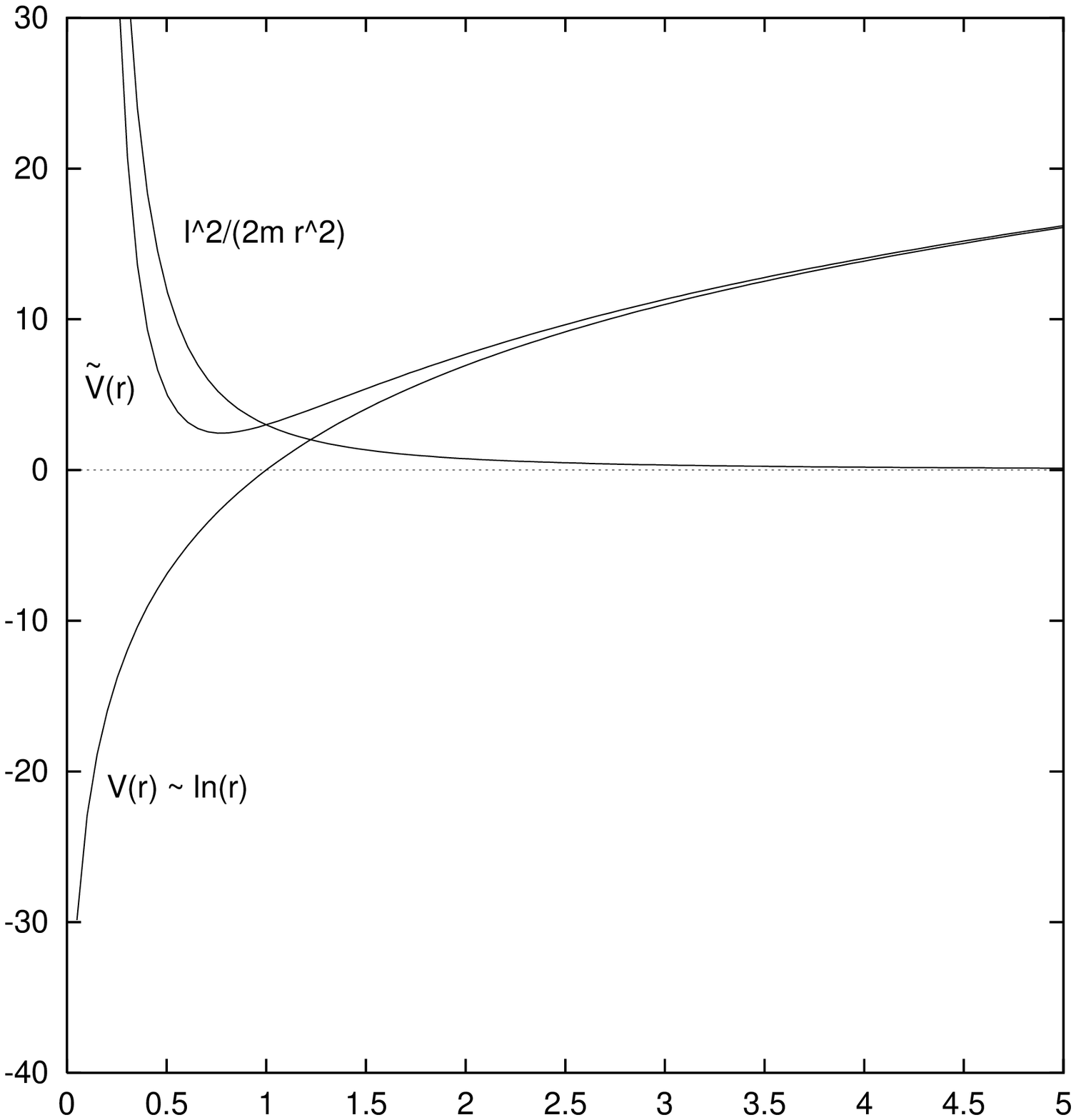}}
\caption{\label{fig:wide} Effective potential for $f(r)\sim - 1/r$ force}
\end{figure}

\subsubsection{$V(r) = a \ln r$}
This corresponds to one over $r$ force $f(r) \sim -a/r$.
The effective potential is given by,
\beq
\tilde{V}(r) = a \ln r + \frac{l^2}{2mr^2}
\eeq

We have the following properties
of $\tilde{V}(r)$
\begin{enumerate}
\item $\lim_{r \rightarrow 0} \tilde{V}(r)  \rightarrow +\infty$.
\item $\lim_{r \rightarrow \infty} \tilde{V}(r)  \rightarrow +\infty$.
\item There is no point of intersection with the $r$ axis, and $\tilde{V}(r)$
is always positive.
\item $\tilde{V}(r)$ reaches a minimum at $r_0= l/\sqrt{am}$. \\
For a minimum point of $\tilde{V}$ we should have
$\partial \tilde{V}/\partial r \mid_{r_0} =0$
and $\partial ^2 \tilde{V}/\partial r^2 \mid_{r_0} > 0$.
\bd
\frac{\partial \tilde{V}}{\partial r} = \frac{a}{r} - \frac{l^2}{mr^3}
\ed
for the point of extremum
\bd
r_0^2 = \frac{l^2}{am}
\ed
The second derivative of $\tilde{V}$ at $r=r_0$,
\beqa
\frac{d^2\tilde{V}}{dr^2} &=& - \frac{a}{r^2} + \frac{3l^2}{mr^4} \\
&=& \frac{1}{r^2} \left(- a + \frac{3 l^2}{mr^2} \right) \\
&=& \frac{1}{r^2} \cdot 2a > 0
\eeqa
Hence it is a point of minimum.
The orbit is always bounded.
\end{enumerate}

\begin{figure}
\resizebox{3.4in}{!}
{\includegraphics*{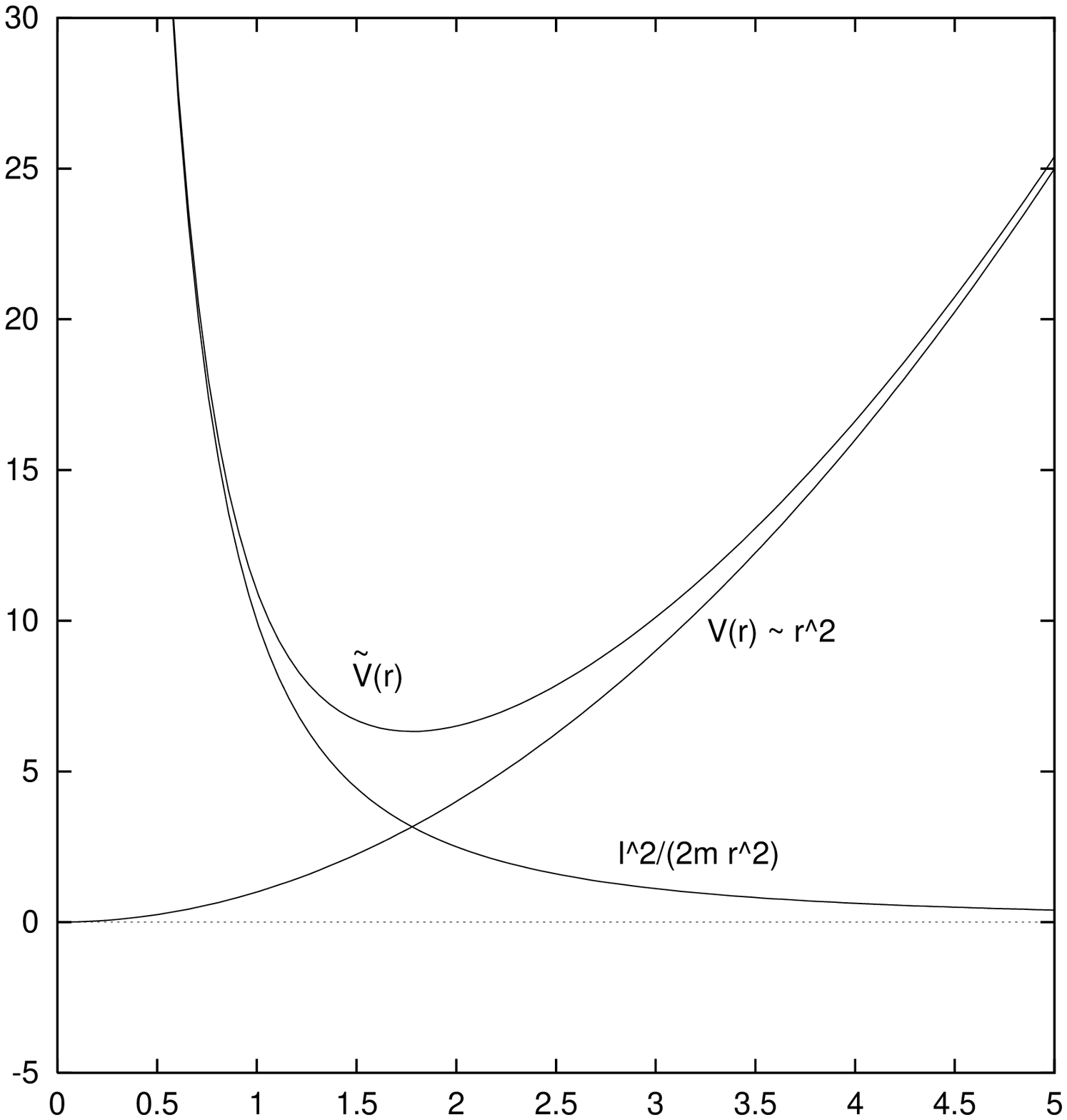}}
\caption{\label{fig:wide} Effective potential for $f(r) \sim - r$ force}
\end{figure}
\subsubsection{$V(r) = a r^n$ where $n>0$}
For this potential the effective potential is given by,
\beq
\tilde{V}(r) = ar^n + \frac{l^2}{2mr^2} \;\;\;\;\;\;\; n >0
\eeq
We have the following properties 
of $\tilde{V}(r)$
\begin{enumerate}
\item $\lim_{r \rightarrow 0} \tilde{V}(r)  \rightarrow +\infty$.
\item $\lim_{r \rightarrow \infty} \tilde{V}(r)  \rightarrow +\infty$.
\item There is no point of intersection with the $r$ axis, and $\tilde{V}(r)$
is always positive.
\item $\tilde{V}(r)$ has a minimum at $r_0$, $r_0^{n+2}=l^2/amn$. \\
For a minimum point of $\tilde{V}$ we should have 
$\partial \tilde{V}/\partial r \mid_{r_0} =0$
and $\partial ^2 \tilde{V}/\partial r^2 \mid_{r_0} > 0$.
\bd
\frac{\partial \tilde{V}}{\partial r} = anr^{n-1} - \frac{l^2}{mr^3}
\ed
for the point of extremum
\bd
{r_0}^{n+2} = \frac{l^2}{amn}
\ed
The second derivative of $\tilde{V}$ with respect to $r$,
\beqa
\frac{d^2\tilde{V}}{dr^2} &=& an(n-1)r^{n-2} + \frac{3l^2}{mr^4} \\
&=& \frac{1}{r^4} \left( an(n-1) r^{n+2} + \frac{3 l^2}{m} \right) \\
&=& \frac{1}{r^4} \frac{l^2}{m}(2+n) >0
\eeqa
\end{enumerate}

Hence it is a point of minimum.
The orbit is always bounded.

The findings for the general power law potential can be
summarized as follows,
\beqar
&& V(r) = sign(n) a r^n \;\;\;\;\;\;\;\;\;\;\; n \neq 0\\
&& f(r) = - abs(n) a r^{n-1}
\eeqar
and
\beqar
&& V(r) = b \ln r \\
&& f(r) = - \frac{b}{r} 
\eeqar

\begin{table}[h]
\caption{Dependence of boundedness on power law.}
\begin{tabular}{|c|l|}
\hline \hline
$n$ of $V(r)\sim a r^n$ & nature of boundedness \\
& \\
\hline
$)-\infty,-2($  & always unbounded \\
& \\
$-2$ & spiralling orbit \\
& \\
$)-2,0($ & bounded or unbounded depending on $E$ \\
& \\
0 ($\ln r$) & always bounded \\
& \\
$)0,+\infty($ & always bounded \\
\hline \hline
\end{tabular}
\end{table}

\subsection{Stable bounded orbit, geometric shape}

The stable bounded orbits in a power law central field can have 
the following forms

\begin{enumerate}
\item $V(r) = -a/r^n, \;\;  0 < n \leq 2$ 
\item $V(r) = b \log r$
\item $V(r) = a \cdot r^n, \;\; n > 0 $
\end{enumerate}
In all the above cases the derivative of $V(r)$ with respect 
to $r$ is always positive, and hence the force is necessarily 
attractive ($f(r) = - \partial V(r)/ \partial r < 0$).

\begin{table}[h]
\caption{Dependence of stability of circular orbits on power law.}
\begin{tabular}{|c|l|}
\hline \hline
$n$ of $V(r)\sim a r^n$ & stability of circular orbits\\
& \\
\hline
$)-\infty,-2)$  & unstable \\
& \\
$)-2,\infty($  & stable \\
\hline \hline
\end{tabular}
\end{table}

In the first case 
the effective one dimensional potential $\tilde{V}(r)$
goes to infinity as $r \rightarrow 0$. $\tilde{V}(r)$ has a minimum
at some $r = r_0$, and it has a negative slope for all $r < r_0$,
and positive slope for all $r > r_0$.
For $E = \tilde{V}(r_0)$ we get stable circular orbit,
and for $\tilde{V}(r_0) < E < 0$.
the orbits are still bounded.
For cases (2) and (3) the circular orbit is stable, and
orbits are always bounded for any energy.

\section{Conclusion}
Existence of bounded orbit for a large class of attractive
central field has been discussed. The generic nature of 
central field bounded orbits is analytically derived.
Certain class of these orbits (Fig. 7) presented in popular 
texts \cite{goldstein}, are shown to be non-feasible.

Small deviation from circularity in the case of
central field is often expressed in terms of inverse of
radial distance ($u=1/r$). 
\beqa
u &=& u_0 + a \cdot \cos(\beta \theta) \\
r &=& \frac{r_0}{1 + a \cdot r_0 \cdot \cos(\beta \theta)}
\eeqa
where $r_0 = 1/u_0$.  
It is interesting to note that these orbits are sometime 
mistakenly identified with diagrams of the form
shown in Fig. 7 \cite{goldfig}. The above expression 
in fact corresponds to a class of orbits that look generically 
like those shown in Fig. 6.

The generic features of central force orbits
discussed here have been verified by computer simulation for 
a large class of central force fields.
The figures shown here (Fig. 2, 3, 5 and 6) 
were generated by these simulations.
Students interested in studying and generating such orbits 
will find Ref.12 helpful.

\section*{Acknowledgement}
The authors wish to express their indebtedness to the well
known texts by Goldstein \cite{goldstein}, Landau \cite{landau}
and Arnold \cite{arnold}. They also acknowledge their
teachers in related graduate courses at Stony Brook,
Prof. Max Dresden, Prof. A. S. Goldhaber and Prof. Leon A. Takhtajan.

Authors gratefully acknowledge the encouragement received
from Prof. Shyamal SenGupta of Presidency College, Calcutta. 
The material presented here was used in a graduate level 
classical mechanics course at Jadavpur University during 1998-2001.
SR wishes to thank his students
for stimulating discussions.


\end{document}